# Hard and Soft Spherical-Bound Stack decoder for MIMO systems

Ghaya Rekaya Ben-Othman, *IEEE Member*, Rym Ouertani, *IEEE Student Member*
and Abdellatif Salah, *IEEE Student Member* *

**Abstract**

Classical ML decoders of MIMO systems like the sphere decoder, the Schnorr-Euchner algorithm, the Fano and the stack decoders suffer of high complexity for high number of antennas and large constellation sizes. We propose in this paper a novel sequential algorithm which combines the stack algorithm search strategy and the sphere decoder search region. The proposed decoder that we call the Spherical-Bound-Stack decoder (SB-Stack) can then be used to resolve lattice and large size constellations decoding with a reduced complexity compared to the classical ML decoders.

The SB-Stack decoder will be further extended to support soft-output detection over linear channels. It will be shown that the soft SB-Stack decoder outperforms other MIMO soft decoders in term of performance and complexity.

## INTRODUCTION

In this work, we are interested in the decoding of multi-antenna systems using spatial multiplexing [1] and linear space time block codes (STBC). In [2], a lattice representation of MIMO systems is proposed. The decoding problem can then be seen as a closest lattice point search problem (CLPS) and the lattice decoders are then used. For MIMO systems, we find in the literature mainly two classes of decoding strategies. On one hand, there are the optimal decoders (ML) such as the sphere decoder and the Schnorr-Euchner algorithm, which have a complexity that increases drastically with the lattice dimension and the constellation size. In the other hand, there are the sub-optimal decoders like the ZF, ZF-DFE,

*Ghaya Rekaya Ben-Othman, Rym Ouertani and Abdellatif Salah are with TELECOM ParisTech, 46 rue Barrault, 75013 Paris FRANCE. Phone: +33 1 45 81 76 33 / + 33 1 45 81 78 40, Fax: +33 1 45 80 40 36, Emails: rekaya, ouertani, salah@telecom-paristech.fr

MMSE and MMSE-DFE decoders which have a low complexity but poor performances. In practice, either performance or complexity can be advantaged. Unfortunately, with these decoders, there is a great margin between performance and complexity and no trade-off is possible.

Sequential decoders are also used to decode MIMO systems, especially in [3] where the authors studied the Fano decoder and a general tree search framework for decoding MIMO systems was established. Another kind of sequential decoders exists in the literature which is the stack decoder introduced by Zigangirov in [4]. Both the stack and the Fano algorithms adopt the same tree search strategy, however their implementation and complexities are quite different.

In [5], the authors focused on the Fano decoder and studied its performances. In this paper, we are interested in the stack one.

We propose here an enhanced stack decoder that we call the Spherical-Bound Stack decoder (SB-Stack). This one constraints the stack search region to a sphere but conserves the original search strategy of the stack algorithm. We will give two versions of the SB-Stack decoder:

- The hard output SB-Stack decoder to decode lattice and finite constellations: we show that the proposed decoder outperforms the classical MIMO ones in term of complexity while keeping ML performances. Furthermore, we show that the SB-Stack decoder can be sub-optimal under some constraints offering performances from ML to ZF-DFE with proportional complexities, which make it possible to get different performance/complexity trade-offs.
- The soft output SB-Stack decoder: we show that the proposed decoder is less complex than the most known soft MIMO ones. It will be observed that the proposed soft SB-Stack offers an interesting improvement of the performances and exhibits a gain up to 2dB.

The paper is composed of three parts. In the first part, we introduce the system model, the notations, an overview of tree search strategies and a description of the sequential algorithms. In the second part, we present the hard output SB-Stack decoder and its adapted version in the case of finite constellations. This latter will be extended to a soft-output decoder in part III and compared to the most known soft algorithms for MIMO systems.

# PART I: AN OVERVIEW OF TREE SEARCH DECODING

## 1. SYSTEM MODEL AND NOTATIONS

### *A. MIMO scheme with spatial multiplexing (SM)*

Let us consider a MIMO system with $M$ transmit and $N$ receive antennas using spatial multiplexing scheme. The channel is assumed to be quasi-static, and the received signal is given by

$$\boldsymbol{y^c_N} = \boldsymbol{H^c_{N\times M}} \cdot \boldsymbol{x^c_M} + \boldsymbol{w^c_N} \tag{1}$$

where $\boldsymbol{H^c}$ is the channel transfer matrix with complex entries $h_{ij}$ representing the fading coefficients between the $i^{th}$ receive and the $j^{th}$ transmit antennas and are modeled by independent Gaussian random variables of zero-mean and variance 0.5 per component. The components of the information transmitted vector $\boldsymbol{x^c}$ are carved in $\mathbb{Z}[i]$ or in a q-QAM constellation and $\boldsymbol{w^c}$ represents the i.i.d complex additive white Gaussian noise vector with zero-mean and variance $\sigma^2$.

A representation of the multi-antenna scheme by a lattice packing was proposed in [6]. This one is obtained by separating the real and imaginary parts as

$$\boldsymbol{y} = \begin{bmatrix} \Re(\boldsymbol{y^c}) \\ \Im(\boldsymbol{y^c}) \end{bmatrix} = \begin{bmatrix} \Re(\boldsymbol{H^c}) & -\Im(\boldsymbol{H^c}) \\ \Im(\boldsymbol{H^c}) & \Re(\boldsymbol{H^c}) \end{bmatrix} \cdot \begin{bmatrix} \Re(\boldsymbol{x^c}) \\ \Im(\boldsymbol{x^c}) \end{bmatrix} + \begin{bmatrix} \Re(\boldsymbol{w^c}) \\ \Im(\boldsymbol{w^c}) \end{bmatrix} \tag{2}$$

$$\boldsymbol{y} = \boldsymbol{H} \cdot \boldsymbol{x} + \boldsymbol{w} \tag{3}$$

$\boldsymbol{H}$ is therefore the equivalent lattice generator matrix of dimension $2N \times 2M$.

### *B. MIMO scheme with STBC*

We now consider a coded system using a linear space time block code [7], such as the TAST codes [5] and the perfect codes [8][9]. The received signal matrix is then given by

$$\boldsymbol{Y^c_{N\times T}} = \boldsymbol{H^c_{N\times M}} \cdot \boldsymbol{C^c_{M\times T}} + \boldsymbol{W^c_{N\times T}} \tag{4}$$

where $\boldsymbol{C^c_{M\times T}}$ represents the codeword matrix. We consider a MIMO symmetric system, *i.e*, $M = N$, and a square code, which means that the temporal code length $T$ is equal to $M$. The lattice representation is obtained here by vectorization and separation of the real and imaginary parts of the received signal $\boldsymbol{Y^c_{N\times T}}$ [8]. The equation (4) becomes therefore

$$\boldsymbol{y^c_{N\cdot T}} = \begin{pmatrix} \boldsymbol{H^c_{N\times M}} & & 0 \\ & \ddots & \\ 0 & & \boldsymbol{H^c_{N\times M}} \end{pmatrix} \cdot \begin{pmatrix} \phi_{11} & \cdots & \phi_{1,M\cdot T} \\ \vdots & \ddots & \vdots \\ \phi_{M\cdot T,1} & \cdots & \phi_{M\cdot T,M\cdot T} \end{pmatrix} \cdot \begin{pmatrix} x_1 \\ \vdots \\ x_{M\cdot T} \end{pmatrix} + \begin{pmatrix} w_1 \\ \vdots \\ w_{N\cdot T} \end{pmatrix}$$

We then get an equivalent system to (1)

$$\begin{aligned} \boldsymbol{y^c_{N\cdot T}} &= \boldsymbol{H^c_{1,N\cdot T\times M\cdot T}} \cdot \boldsymbol{\phi^c_{M\cdot T\times M\cdot T}} \cdot \boldsymbol{x^c_{M\cdot T}} + \boldsymbol{w^c_{N\cdot T}} \\ &= \boldsymbol{H'^c_{N\cdot T\times M\cdot T}} \cdot \boldsymbol{x^c_{M\cdot T}} + \boldsymbol{w^c_{N\cdot T}} \end{aligned} \tag{5}$$

Then, the separation of the real and imaginary parts is applied on the former equation as in (2), and the coded system is therefore given by

$$\begin{aligned} \boldsymbol{y_n} &= \boldsymbol{H'_{n\times n}} \cdot \boldsymbol{x_n} + \boldsymbol{w_n} \\ &= \boldsymbol{H} \cdot \boldsymbol{x} + \boldsymbol{w} \end{aligned} \tag{6}$$

where we define by $\boldsymbol{H} = \boldsymbol{H'_{n\times n}}$ the equivalent lattice generator matrix of dimension $n = 2M^2$.

### *C. Decoding scheme*

Under the assumption of a perfect knowledge of the channel state information at the receiver, the ML decoding rule is given by

$$\hat{\boldsymbol{x}} = arg\left(\min_{x\in\mathbb{Z}^n\, or\, x\in(q-QAM)^n} \|y - \boldsymbol{H}\cdot\boldsymbol{x}\|^2\right) \tag{7}$$

The obtained system model can then be decoded by either sub-optimal decoders like the ZF, ZF-DFE, MMSE decoders [10][11], or optimal ones such as the sphere decoder and the Schnorr-Euchner algorithm [12][13] which are basically tree-search algorithms.

More generally, to apply tree-search algorithms, we need first to expose the tree structure. A QR or a Cholesky decomposition can be then applied on the lattice generator matrix $\boldsymbol{H}$. These two methods are quite equivalent, however the QR is more complex than the Cholesky decomposition but it is more stable numerically [14]. For our simulations, we then choose to apply the QR decomposition. Then, the

system can be written as

$$\boldsymbol{y} = \boldsymbol{Q} \cdot \boldsymbol{R} \cdot \boldsymbol{x} + \boldsymbol{w} \tag{8}$$

where $\boldsymbol{Q}$ is an orthogonal matrix and $\boldsymbol{R}$ an upper triangular one. The multiplication of both sides of (8) by the transpose of $\boldsymbol{Q}$ does not change the decoding problem and we get

$$\boldsymbol{z} = \boldsymbol{Q}^{\dagger} \cdot \boldsymbol{y} = \boldsymbol{R} \cdot \boldsymbol{x} + \boldsymbol{Q}^{\dagger} \cdot \boldsymbol{w}$$

Exploiting the upper triangular form of $R$, one can solve the decoding problem using a tree search algorithm and the ML metric to be minimized is then given by

$$\|\boldsymbol{z} - \boldsymbol{R} \cdot \boldsymbol{x}\|^2 \quad / \quad x \in \mathbb{Z}^n \, or \, x \in (q - QAM)^n \tag{9}$$

Throughout this paper, we consider a tree rooted at a fictive node $x_{root}$. The node at level $k$ is denoted by the vector $x^{(k)} = (x_n, x_{n-1}, ..., x_k)$ where $x_j, j = 1, \ldots, n$ are the components of $\boldsymbol{x}$. Moreover, the branches of the tree at level $k$ define all the possible values that can be taken by $x_k$, and each node $x^{(k)}$ is associated with the squared distance

$$f(x^{(k)}) = \sum_{i=k}^{n} f_i(x_i) \tag{10}$$

where $f_i(x_i) = \left| z_i - \sum_{j=i}^{n} r_{i,j} x_j \right|^2$. We call $f(x^{(k)})$ the cost function of the node $x^{(k)}$. It represents the ”sub-distance” between the received and the transmitted signal at the level $k$.

The tree search consists in exploring the tree nodes in order to find the path $(x_n, x_{n-1}, ..., x_1)$ with the least cost. In the literature, we find different tree search strategies. In the next section, we will present the most known ones.

## 2. An overview of tree search strategies

### *A. Breadth First Search (BrFS)*

The breadth first search algorithm is a tree search algorithm that starts from a root node $x_{root}$ and explores all its neighboring. Then, for each of these nodes, it explores all their unexplored neighbors, and so on until it hits the end of the tree. The BrFS is then an exhaustive tree search algorithm. It moves from level $k+1$ to the level $k$ until it explores all the nodes in the first one. The solution found is then the ML one.

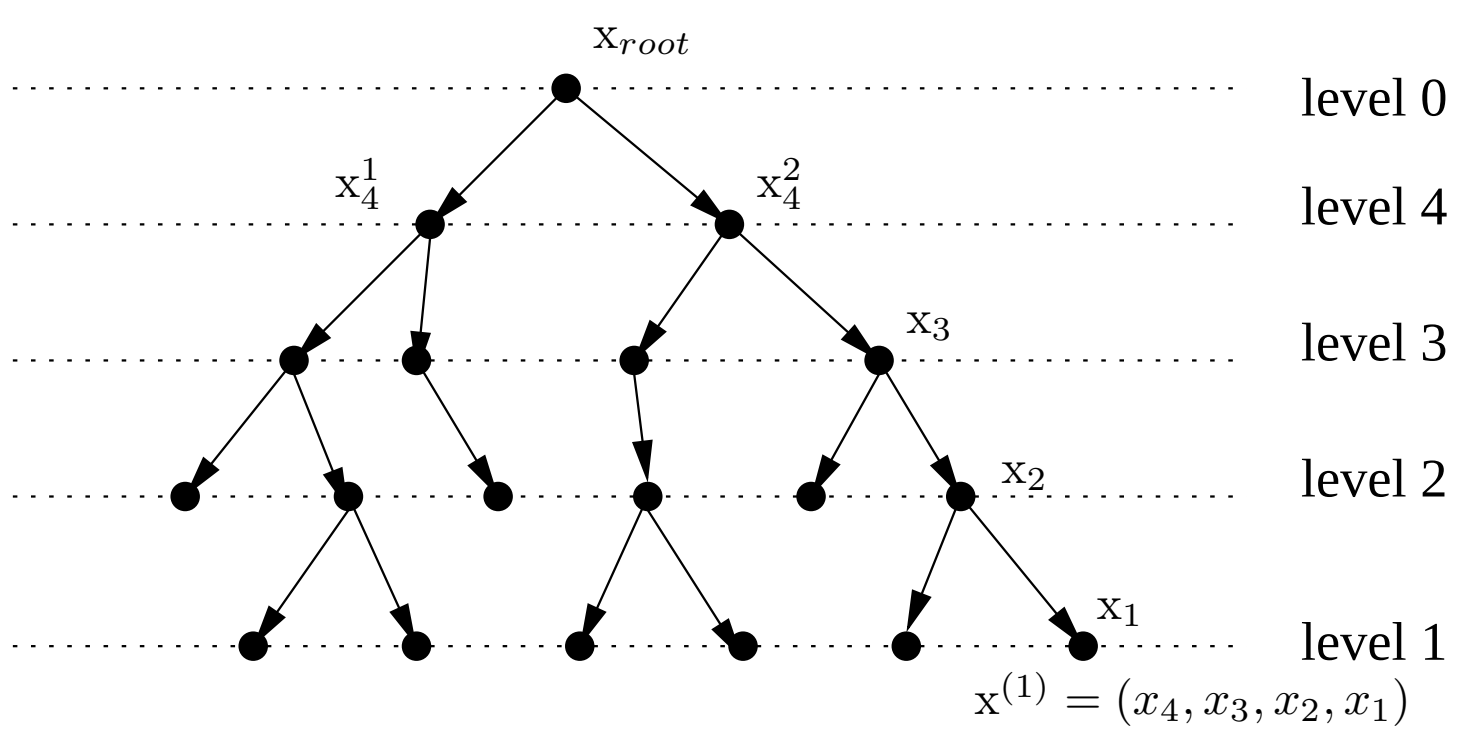

Fig. 1. Example of a tree structure with a dimension of n=4

### *B. Depth First Search (DFS)*

Unlike the BrFS, the DFS algorithm starts from the root node, explores its first child node and proceeds by going deeper and deeper until the end of the tree or until it hits a node that has no children. Then, the algorithm backtracks and returns to the most recent node being expanded. We note that this algorithm needs more memory to 'remember' which nodes having been already visited. However, since it explores all the possible paths in the tree, the DFS is an exact-ML algorithm.

### *C. Best First Search (BeFS)*

One can see the BeFS as an optimization of the BrFS. In fact, the BeFS strategy aims to find the best path in the tree by expanding only the most promising nodes chosen according to some rule. In general, the BeFS uses an *evaluation function* and selects the next node to expand with the best score. In fact, starting from a given node, the algorithm evaluates first all its successors and selects the one to expand with the best score and so on until finding the final node.

### *D. Branch and Bound algorithm*

Visiting all the tree nodes to find the one with the shortest path, using one of the three strategies described above, is prohibitively complex. However, this complexity can be reduced using the Branch and Bound algorithm (BB) which comes to establish constraints on the tree search by using a *bounding* *function*. This means that the algorithm chooses the nodes to expand by comparing their scores against this function. If the cost node is within the defined bounders, the node will be explored, else the node will be jumped, which allows to limit the expanding of some unnecessary nodes and advantages the most promising ones.

The sphere decoder and the Schnorr-Euchner algorithm are both BB algorithms using a depth-first-search strategy. In fact, they start from the upper level in the tree and first consider one possible value $\tilde{x}_n$ inside the bounded region and conducts a depth search over the sub-trees $\{\boldsymbol{x}/x_n = \tilde{x}_n\}$ before going back to another sub-trees $\{\boldsymbol{x}/x_n \neq \tilde{x}_n\}$, and so on.

The sequential decoders also conducts BB algorithms using a BeFS strategy. In the following, we will focus on the most known ones, namely the Fano and the stack algorithms. A description of these sequential decoding algorithms is therefore given.

## 3. Sequential decoders

The sequential decoders were originally proposed to decode binary trellis codes[15]. The most used one is the Fano decoder introduced in 1963[16]. Later, Zigangirov proposed in 1966 a sequential algorithm using a stack storage (or memory). In the 1960s, memory allocation represented an additional problem, that's why the Fano decoder was more suitable for hardware implementation and so far the stack decoder has not been widely used. Nowadays, the price of memories is continually dropping and the stack decoder is therefore being of great interest.

In [3], the authors have rediscovered the Fano decoder and applied it to decode MIMO schemes. In this work, we will focus on the stack decoder and bring the necessary modifications to decode lattice and finite constellations. Before presenting the modified stack decoder, we recall the principal of the original Fano and stack ones.

### *A. Fano decoder*

We will detail the search strategy of the Fano decoder. Let us suppose that the decoder is at a some node $x_k$ of a level $k$ in the tree. The decoder can choose between 3 options: proceed forward to the next node at level $k-1$, move back to a predecessor node $x_{k+1}$, or move laterally to a neighbor node at the same level $k$. We call $f_{look}$ and $f_{pre}$ the costs of respectively the successor and the predecessor node. At each step, the decoder looks forward to the best next node. The best node is the one having the least metric. The decoder can visit a node if its metric is smaller than a certain threshold $\Upsilon$. Each time a new node is visited, $\Upsilon$ is lowered by a step-size $\Delta$. If the successor's cost is larger than the current threshold, the decoder looks at the predecessor node. If $f_{pre} \leq \Upsilon$, the decoder moves backward to this node and looks forward to its next best successor if it exists. Otherwise, that means that all the nodes which costs are smaller than the threshold are explored. The decoder increases then $\Upsilon$, tries to move forward to the next node, and so on. The algorithm terminates when a leaf node is reached. We note that, from a current node, the Fano algorithm moves either to its predecessor or to its successor, but never jumps. However, since the decoder requires no memory, it may return to the same node several times which includes additional computations needed to decode a given sequence.

### *B. Stack decoder*

For the stack algorithm, the search principle and strategy remain the same. However, the main difference is that the stack stores all the paths crossed by the algorithm in an ordered list called ”stack” or ”memory”, whereas the Fano only retains the best path. In fact, starting from a root node $x_{root}$, the stack algorithm generates all the child nodes of $x_{root}$. We call a child node the successor node. The algorithm computes then the respective costs of those nodes according to the cost function in equation (10) and stores them in an increasing order in the memory, so that the top node of the stack is the one having the least cost. Afterwards, the algorithm generates the children of the top, computes their costs, places them in the memory and removes the top node being just extended. The algorithm reorders again the stack, generates the child nodes of the current top node, and so on. The algorithm terminates when a path of length $n$ is found on the top of the stack, in other words, when a leaf node reaches the top of the list.

The flowchart of the stack algorithm is presented in Fig.2, where we define by $Gen(x^{(k)})$ the function that generates all the children nodes of $x^{(k)}$ and by $Sort()$ the function that reorders the stored nodes in the stack.

Note that, since the stack decoder requires a memory, it visits fewer nodes than the Fano which may return to a node having already been visited. Consequently, the stack decoder is much faster and so less complex than the Fano. In the sequel, we will propose a modified stack decoder that is less complex than the original one.

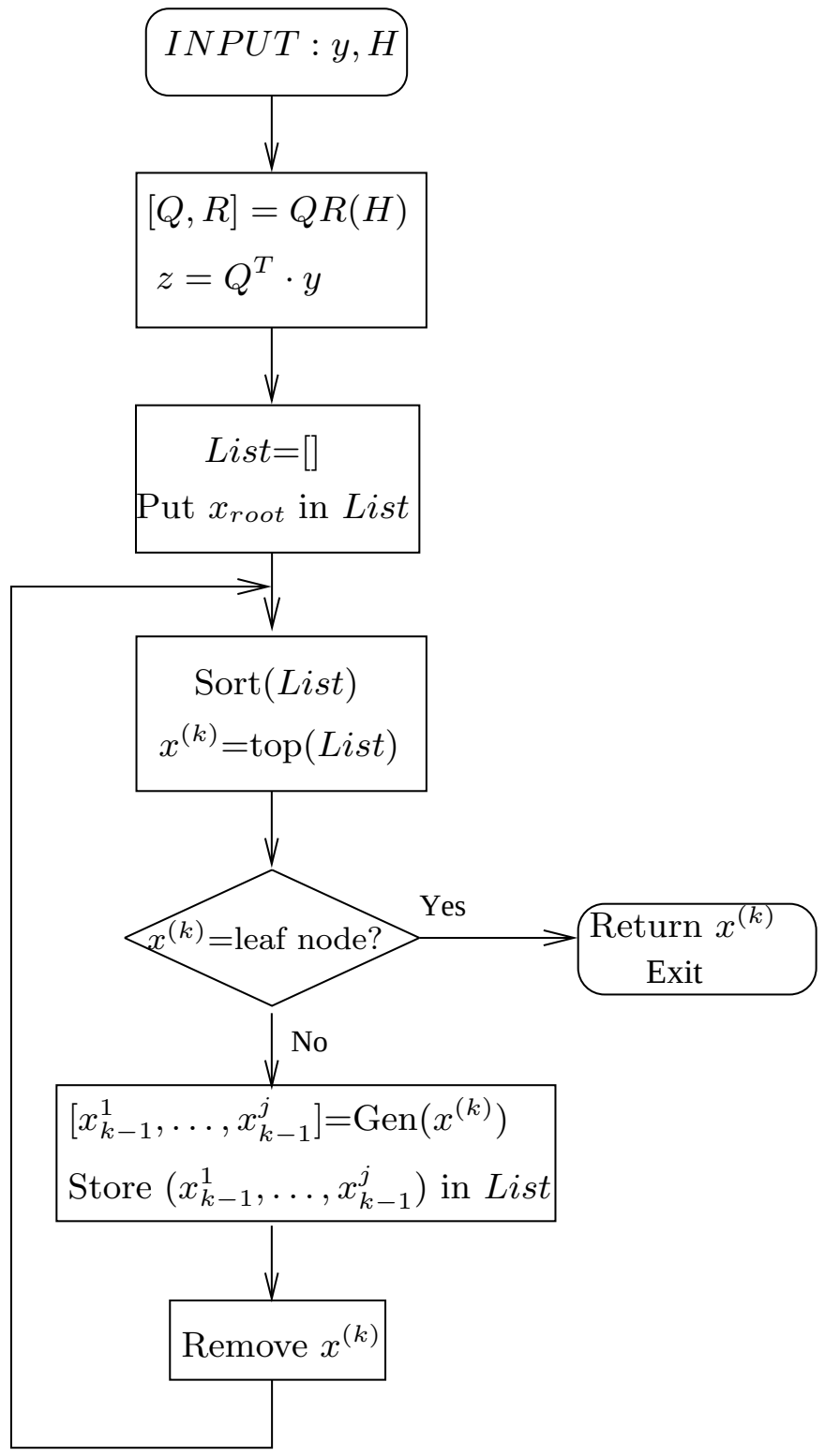

Fig. 2. Flowchart of the stack decoder

# Part II: Hard decoding using the stack algorithm

## 1. Lattice decoding

Stack decoding was originally designed to decode binary trellis codes, where the codeword is taken in a finite alphabet. However, considering a lattice, the codeword is taken in the infinite field $\mathbb{Z}^n$ which leads to an infinite tree structure. Applying the stack decoder seems to be impossible in this case. Our purpose is then to propose a modified version of the stack algorithm in order to decode lattice.

### *A. $1^{st}$ approach*

To apply the stack decoder, we should search for the closest point in a finite region $\Lambda \subset \mathbb{Z}^n$. Unfortunately, the truncation of the tree will affect the performances. In fact, if the transmitted codeword belongs to $\Lambda$, the decoder will systematically find it, however an error is occurred when the codeword is out of the search region. The main challenge is then how to choose the optimal region $\Lambda$.

Yet, the triangular form of the lattice basis reminds us the Schnorr Euchner enumeration strategy[17]. The key idea of this algorithm is to view the lattice as a superposition of $n$ hyperplanes and to start the search by projecting the received vector on the nearest hyperplane. The resulting point is then recursively projected on the following $n-1$ hyperplanes. The point found is the ”babai point” and it corresponds to the ZF-DFE point[6]. Similarly to the Schnorr Euchner, the proposed search algorithm is based on the babai point, however the search strategy and the construction of the tree are quite different.

In fact, the Schnorr Euchner consists in enumerating all the possible nodes inside a bounded region by zigzagging around the babai point using a DFS strategy. We propose here an algorithm that constructs a tree centered at the babai point $\boldsymbol{u}$. At each level, it enumerates the neighbor lattice points defined as $\boldsymbol{u} \pm \boldsymbol{t} = (u_1 \pm t_1, u_2 \pm t_2, ..., u_n \pm t_n)$ where $\boldsymbol{t}$ is a vector in $\mathbb{Z}^n$, a BeFS strategy is further applied on this tree.

Applying this algorithm, we can delimit the size of the constructed tree by choosing the number of the neighboring lattice nodes of the babai point that we would consider. However, the ML point is not guaranteed to be included in the considered tree. To reach it, we should enlarge the search region. Meanwhile, that implies to have a denser tree which leads to a more complex decoding task.

In Fig.3, we plot the symbol error rate as a function of the signal to noise ratio (SNR) given in dB scale, for a $4 \times 4$ MIMO system using a SM. First, we proceed by considering the search region defined as $\Lambda_a = \{u_i - 1, u_i, u_i + 1, i = 1, \ldots, n\}$. This induces that the lattice points concerned with the search algorithm are only the immediate neighbors of the babai point. Nevertheless, in bad channel conditions, the ML point may be far-off and then unreachable in this case. This is shown in the curve (a), where the performances are sub-optimal and exhibit a loss of 2dB from ML. For the same system, we have progressively enlarged the search region and observed the algorithm’s behavior. The curves (a)-(d) report the performances obtained by respectively considering the search regions $\Lambda_a = \{u_i - 1, u_i, u_i + 1\}$, $\Lambda_b = \Lambda_a \bigcup \{u_i - 2, u_i + 2\}$, $\Lambda_c = \Lambda_b \bigcup \{u_i - 3, u_i + 3\}$ and $\Lambda_d = \Lambda_c \bigcup \{u_i - 4, u_i + 4\}$. As shown in Fig.3, the decoder provides sub-optimal performances, but it approaches the ML as well as we cover a larger search region. However, the complexity increases with the performances. Therefore, a compromise may be done and this decoding algorithm can be of great interest. In fact, at the start of the algorithm, one only needs to choose the performance-complexity trade-off to reach, to define the appropriate search region.

In simulations, we have considered a uniform vector $\boldsymbol{t}$. One can think to use a vector $\boldsymbol{t}$ with large $t_i$ for first components and small $t_i$ for the last ones. This choice may be efficient due to the problem of error propagation in the tree search.

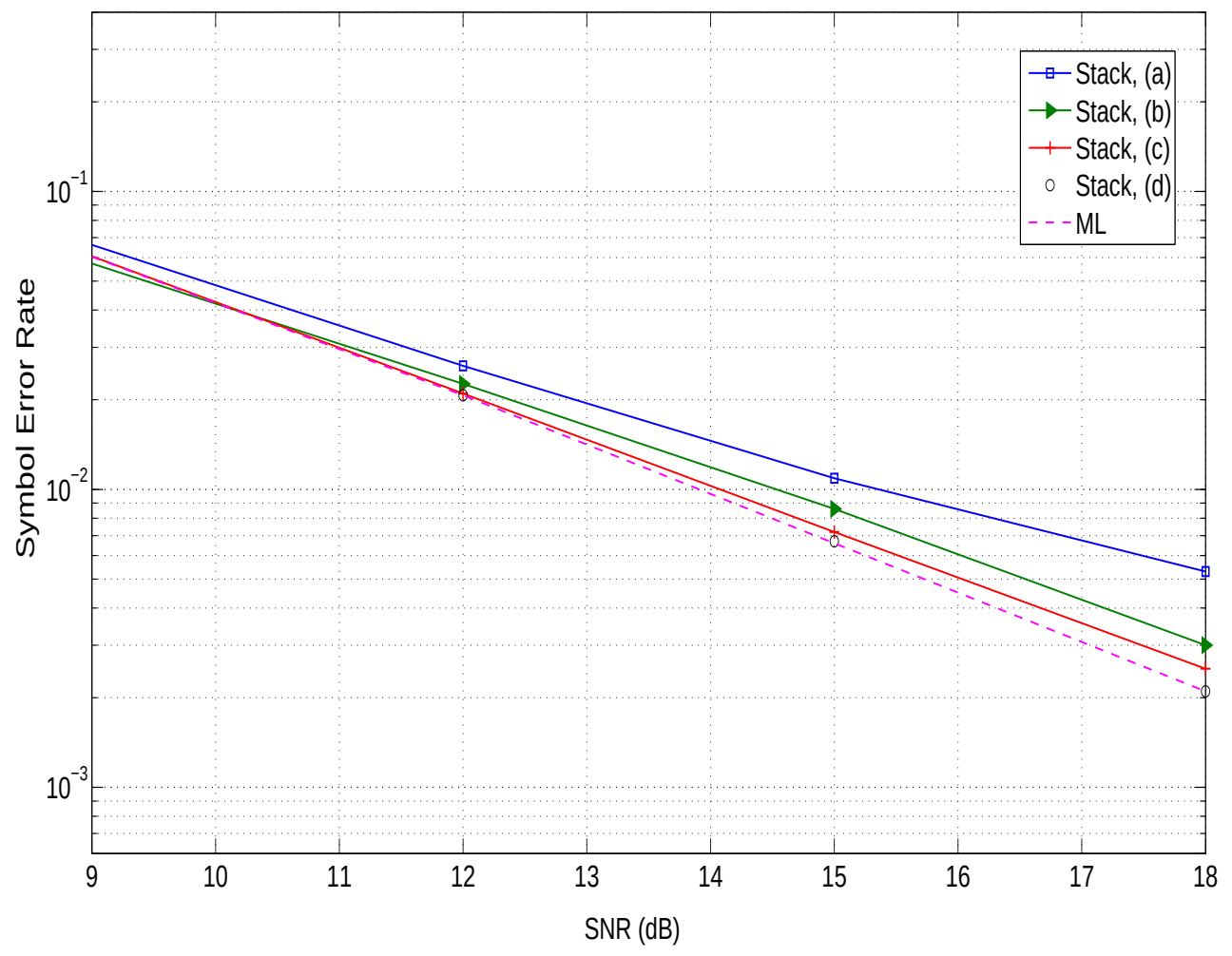

Fig. 3. Performances of a MIMO system using a SM with $M = N = 4$, obtained for different search regions

### *B.* $2^{nd}$*approach (SB-Stack decoder)*

In the $1^{st}$ approach, the search region is centered on the babai point. However, this latter is generally a rough estimation of the transmitted codeword, then centering the search region on it is not the best way to do since the ML solution may not be reached, as shown in Fig.4.

Therefore, we propose here a second approach for the lattice decoding inspired by the sphere decoder. The principle of this latter is to enumerate all the lattice points found in a sphere of a radius $C$ centered on the received point. Each time a point is found, the radius is updated, which limits the number of the enumerated points but also ensures the closest point criterion. The sphere decoder uses the DFS strategy.

Like the sphere decoder, the SB-Stack algorithm only explores the lattice points inside the sphere with the radius $C$ using the BeFS strategy, which induces to the definition of an upper and a lower bounds for each lattice point component (the tree nodes). Starting from the root node, the algorithm then computes the upper and lower bounds of the first component $x_n$, respectively denoted $b_{inf,n}$ and $b_{sup,n}$ and generates all the nodes within these bounds, which represent the branches of the tree at level $n$.

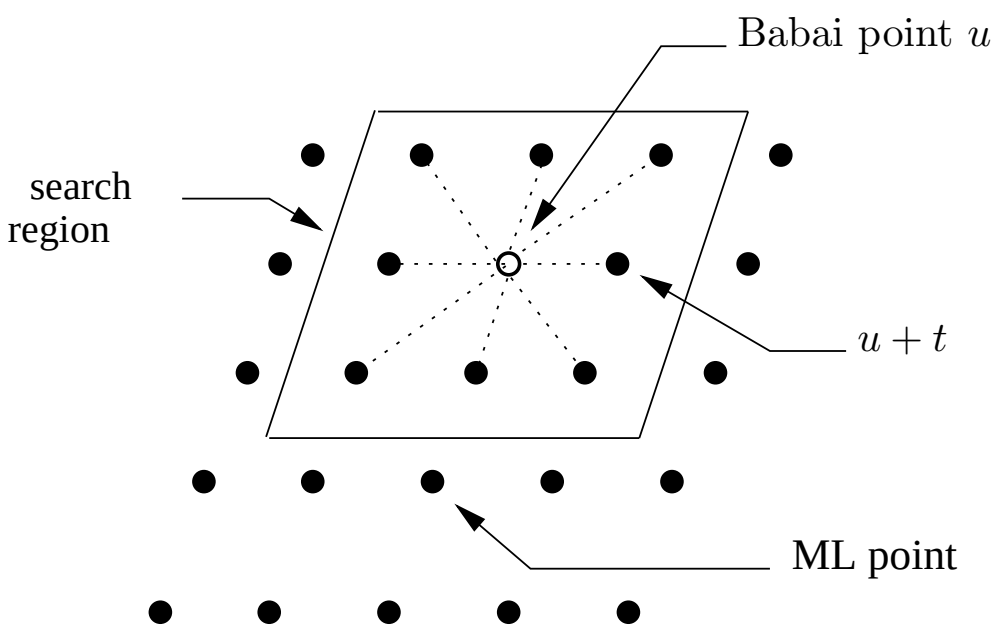

Fig. 4. Example of a lattice defined in $\mathbb{Z}^2$, the search region does not contain the ML point

The algorithm then computes the cost of each node $x_n$ and stores them with their respective costs in the stack memory. After that, it reorders the nodes in the memory in an increasing order according to their costs, selects the top node, then computes the bounds of the next level $(n-1)$ using the value of the top node $x_n$, generates all its possible children and stores them in the memory after removing the top node. We call children of $x_n$ all the nodes $x_{n-1}$ within the bounds $b_{inf,n-1}$ and $b_{sup,n-1}$. This procedure is repeated until a leaf node reaches the top of the memory.

Note that, although the apparent similarity between the traditional sphere decoder and the SB-Stack algorithm, these two search algorithms raise great differences. In this way, unlike the sphere decoder the radius in the SB-Stack decoder remains unchanged during all the decoding process, while for the sphere decoder the radius is being updated each time a point is found. In fact, the sphere decoder proceeds by first searching one candidate solution by performing a DFS. This candidate is then an estimation of the closest solution, which means that, the ML point is at least from this distance to the received point, and it is then unnecessary to look for candidates beyond this metric. Consequently, the sphere radius can be reduced to this new distance.

However, the search strategy for the stack algorithm is different. As described above, at each step the algorithm may backtrack to a higher node before reaching the end of the tree which corresponds to a candidate solution. Thus, there is no possibility to evaluate the ML distance in progress of the algorithm, and the radius must therefore be fixed.

For more clarity, we will now detail the bounds calculation which is the same as those of the sphere decoder in [12].

*Bounds calculation:*

First, remind that the distance to minimize is given by $\|\boldsymbol{z}-\boldsymbol{R}\cdot\boldsymbol{x}\|^2$. Let us write $\boldsymbol{z}=\boldsymbol{R}\cdot\boldsymbol{\rho}$, where $\boldsymbol{\rho}$ is the ZF point. It represents the coordinate of the vector $\boldsymbol{z}$ in the new lattice generated by $\boldsymbol{R}$. The euclidean distance is now written in the lattice system as:$\|\boldsymbol{R}\cdot(\boldsymbol{\rho}-\boldsymbol{x})\|^2 = \|\boldsymbol{R}\cdot\boldsymbol{\xi}\|^2$, where $\boldsymbol{\xi}$ defines the coordinate of the translated point $\boldsymbol{x}$.

In this case, the lattice points considered in the metric minimization are those within the distance $C$ from the received point $\boldsymbol{z}$, we then have

$$\|\boldsymbol{R}\cdot\boldsymbol{\xi}\|^2=\sum_{i=1}^{n}\left(\sum_{j=i}^{n} r_{ij}\xi_j\right)^2 \leq C^2 \tag{11}$$

Let now define by: $q_{ii}^1 = r_{ii}^2$ for $i=1,\ldots,n$, and $q_{ij}^1=\frac{r_{ij}}{r_{ii}}$ $,i=1,\ldots,n$ for $i=1,\ldots,n$, and $j=i+1,\ldots,n$. The equation (11) is rewritten as

$$\|\boldsymbol{R}\cdot\boldsymbol{\xi}\|^2=\sum_{i=1}^{n} q_{ii}^1\left(\xi_i+\sum_{j=i+1}^{n} q_{ij}^1\xi_j\right)^2 \leq C^2 \tag{12}$$

By working backward, we define the bounds at any level $i$ by

$$b_{inf,i}=\left\lceil -\sqrt{\frac{T_i}{q_{ii}^1}}+S_i\right\rceil \leq x_i \leq \left\lfloor \sqrt{\frac{T_i}{q_{ii}^1}}+S_i\right\rfloor = b_{sup,i} \tag{13}$$

where we refer to $T_i$ and $S_i$ by

$$\begin{aligned} T_{i-1} &= C^2-\sum_{l=i}^{n} q_{ll}^1\left(\xi_l+\sum_{l=i}^{n} q_{lj}^1\xi_j\right)^2 \\ S_i &= \rho_i+\sum_{l=i+1}^{n} q_{il}^1\xi_l \end{aligned} \tag{14}$$

The flowchart of the SB-Stack decoder is given in Fig.5.

Note that the bounds of a given node $x_i$ are dependent of those of the former node $x_{i+1}$ as shown in the equation (14). Therefore, the SB-Stack algorithm also stores the bounds of each generated node, that will be used to compute its children nodes when this latter should be extended. Consequently, the SB-Stack stores more information than the classical stack algorithm and the price to pay is an increasing memory size. But the search region limitation allows to visit fewer nodes and so to converge more quickly and ensures to obtain the ML solution.

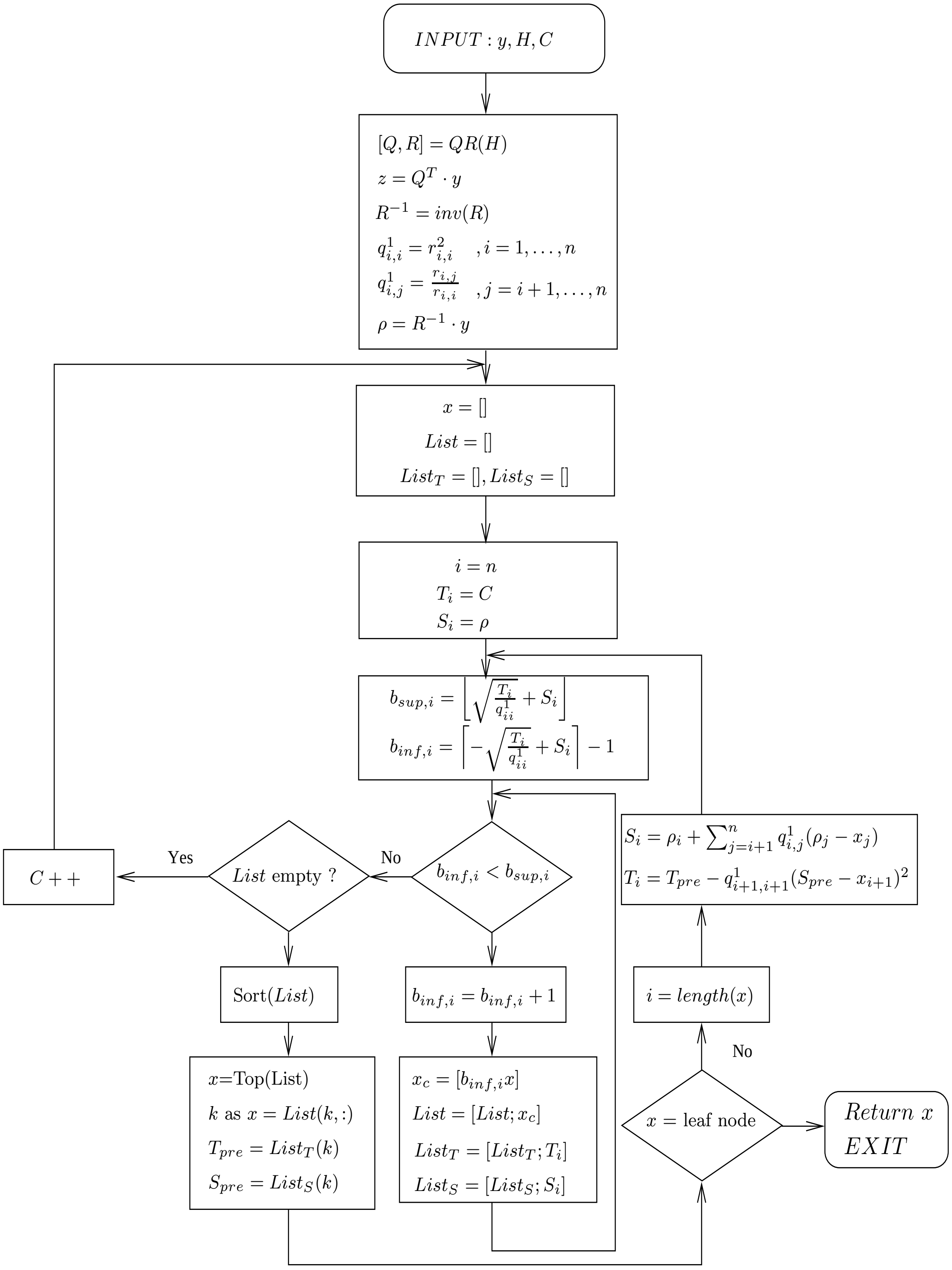

Fig. 5. Flowchart of the Spherical-Bound-Stack decoder

Once the bounds are computed, the algorithm generates all the nodes corresponding to the different values of $x_i$. If there is no valid value of $x_i$ (i.e. $b_{inf,i} > b_{sup,i}$), that means that $x_{i+1}$ has no children and thus this intermediate node has no chance to yield to a leaf node. It will be then removed from the stack list. However if the stack is empty, that means that no lattice point was found inside the sphere. In this case, the sphere radius must be increased and the algorithm is then restarted.

From this observation, it is clear that the complexity of the algorithm depends also on the sphere radius. In fact, if $C$ is too large, we obtain too many points, and so a large tree search. But if $C$ is too small, we obtain no points. Furthermore, for small SNR, the received signal is much affected by the noise and a large radius is needed, while for high SNR, the ML point is close to the received signal and a small radius is sufficient. In [18], a formula to choose the optimal radius as a function of the SNR was first reported by Hassibi and *al.* as

$$C^2 = 2.n.\sigma^2 \tag{15}$$

In the other hand, the lattice matrix $\boldsymbol{H}$ is the product of the fading matrix $\boldsymbol{H_1}$ and the unitary code generator matrix $\phi$(equation (5)). Consequently, in presence of a deep fading the lattice can be much distorted and may be more stretched from some axes than others. It is then more suitable to compute the sphere radius according to the fading too. Therefore, a formula taken into account both the noise variance and the fading matrix was proposed in [19]

$$C^2 = \min(2.n.\sigma^2, \min(diag(\boldsymbol{H^T} \cdot \boldsymbol{H}))) \tag{16}$$

As both the sphere and the SB-Stack decoders are ML, we will focus our comparison on the complexity which we count as the total number of multiplications of the search phase. In Fig.6, we plot the complexity as a function of the SNR for a $2\times2$ and a $4\times4$ MIMO systems using SM. We only consider the complexity of the search phase as the same pre-processing phase (QR decomposition) is made for both decoders.

We can see that the SB-Stack offers a considerable complexity reduction for different lattice dimensions, which is about 40% than the sphere decoder for the $2 \times 2$ system and 50% for the $4 \times 4$ one at low SNR. This important complexity reduction is due to the search strategy of the SB-Stack decoder which allows to look inside the sphere only for the most promised points, unlike the sphere decoder which checks all the lattice points inside the sphere.

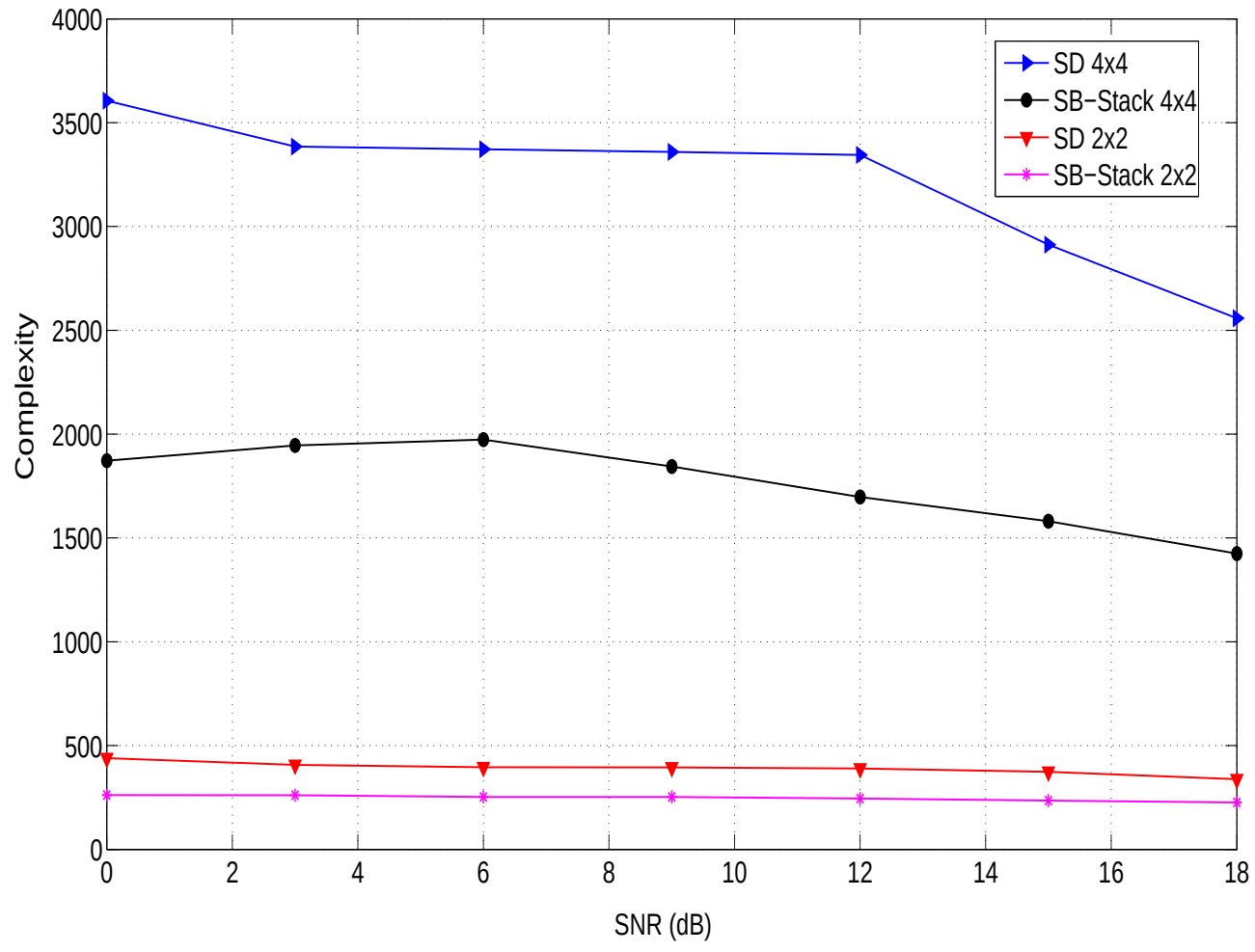

Fig. 6. Performance and complexity of the SB-Stack decoder for a $2 \times 2$ and a $4 \times 4$ MIMO systems using a SM

In practical transmission schemes, we do not consider information symbols in $\mathbb{Z}^n$ but in finite constellations. The most used ones for MIMO schemes are the $QAM$ constellations. We propose in the sequel to modify the stack and the SB-stack decoders to take into account the finite $QAM$ constellations.

## 2. Decoding finite constellations

In this section, we will focus on the decoding of finite constellations using the stack decoder. As in section II.1, we propose two approaches: the first approach is largely inspired by the original stack decoder, while the second one is a readjustment of the proposed SB-Stack algorithm described above.

### *A.* $1^{st}$*approach*

Using finite constellations tasks the decoding problem in its original context where the stack decoder was applied to decode binary codes. In our case, the tree is no longer binary even though it remains finite. As a first and natural approach, we propose to use the original stack decoder, but instead of the binary values of the tree nodes, we consider the correspondent interval of the $q-QAM$ constellation. For example, for a $16-QAM$ constellation, each tree node belongs to the set $I_c = \{\pm 1, \pm 3\}$, then we have a $4-ary$ tree. More generally, for a $q-QAM$, the nodes to consider are in $\left\{\pm 1, \pm 3, \ldots, \pm \sqrt{q}-1\right\}$. Consequently, the tree structure is directly linked to the used constellation, and for large sizes, the information set to which belong the symbols is too large which leads to an excessively dense tree and thus to a very complex decoding. To illustrate this, we represent in Fig.7 the complexity of the stack decoder for 16 and $64-QAM$ constellations as a function of SNR for a $4 \times 4$ MIMO system using SM. As we can see, the complexity increases as the constellation size increases. We also note that, for large sizes, the stack decoder is much more complex than the sphere decoder. In fact, for each level the stack decoder generates all the possible nodes of the constellation, while the sphere decoder only selects the closer ones. Moreover, this complexity is especially high for low SNR where the decoder crosses more nodes to reach the optimal solution.

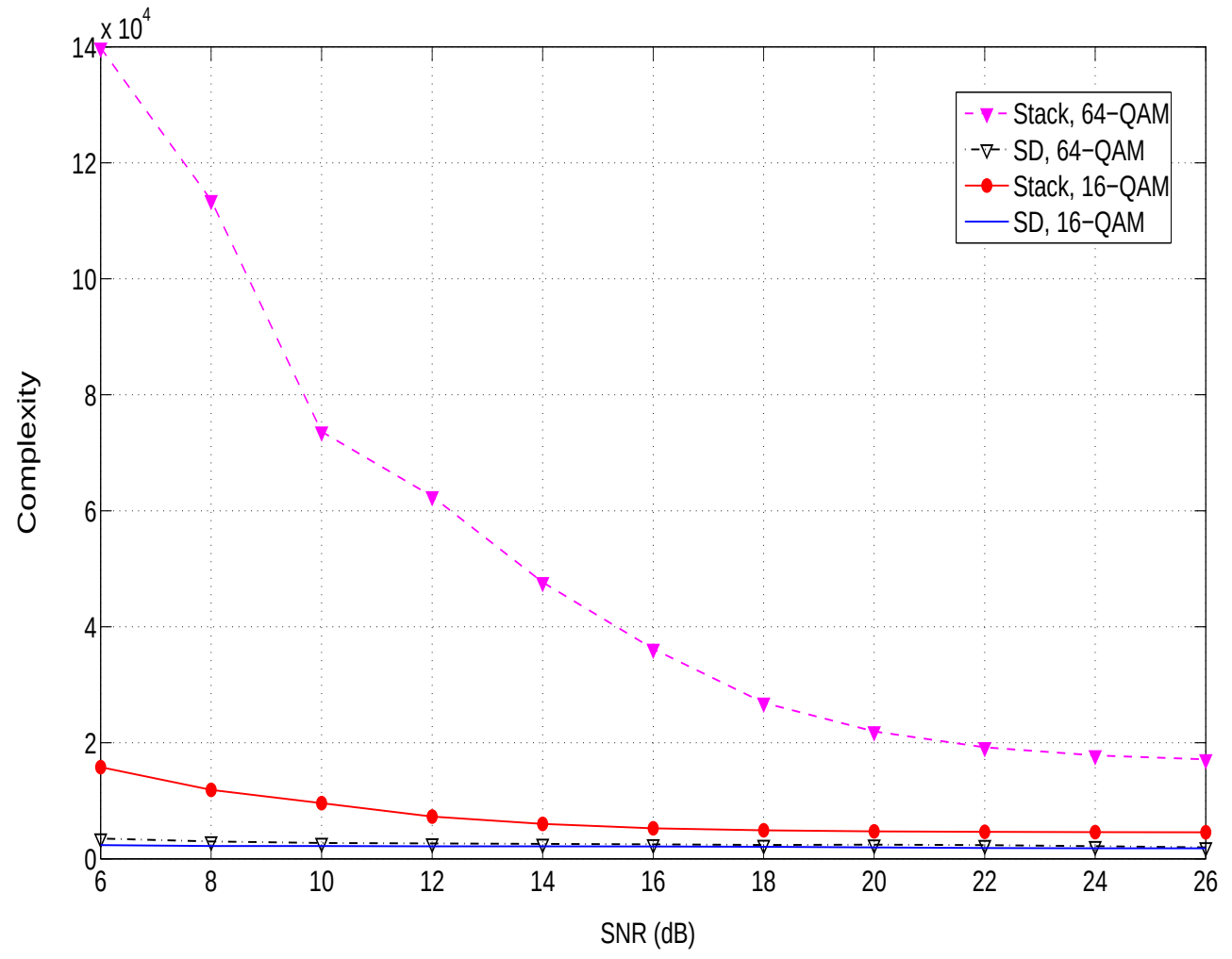

Fig. 7. Comparison of the complexity of the stack decoder and the sphere decoder using different $QAM$ constellations, for a MIMO system with SM, $M = N = 4$

The complexity therefore comes from the high number of visited nodes. In [20], a tree search algorithm was proposed. This one performs a stack algorithm but it limits the size of the stack so that it only retains K nodes at each level of the tree. Nevertheless, the nodes selected in the stack may not lead to the shortest path but those corresponding to the smallest metric may have been discarded in higher levels in the tree. Consequently, the ML solution is not reached. So, this algorithm allows to reduce the complexity by limiting the number of the generated nodes but at the price of a performance loss.

We propose in the sequel to use the SB-Stack decoder described in section II.1. We then give in the following the necessary modifications to adapt it to decode finite constellations.

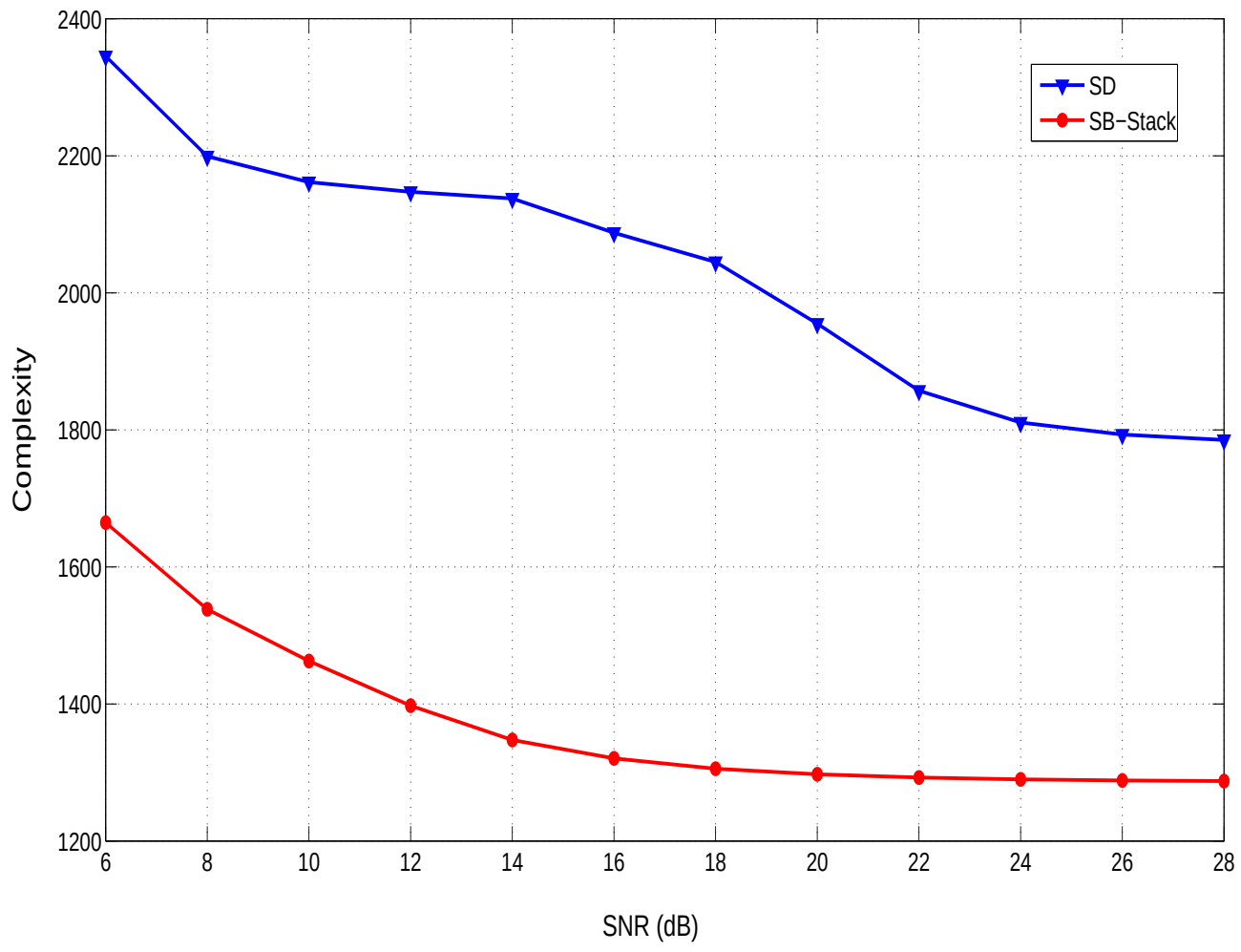

Fig. 8. Complexity of the SB-Stack decoder *vs* the sphere decoder for a $4 \times 4$ MIMO system with SM, using a $16 - QAM$ constellation

### B. $2^{nd}$ approach using the SB-Stack algorithm

The SB-Stack algorithm as presented above is expected to decode lattice. For information symbols taken in a $q - QAM$ constellation, each component of $\boldsymbol{x}$ belongs to the finite interval $I_c = \left[\pm 1, \pm 3, \ldots, \pm\sqrt{q} - 1\right] \subset \mathbb{Z}^n$. The nodes concerned with the search algorithm inside the sphere are only the ones that belong to the constellation and so those which are within $I_c$. Furthermore, in order to restrict the search to a set of $\mathbb{Z}^n$, we consider the transformation $u_i = (x_i + \sqrt{q} - 1)/2$. The new bounders of the constellation are then given by $I_{c,\mathbb{Z}} = \left[0, 1, 2, \ldots, \sqrt{q} - 1\right]$. Consequently, the nodes that we look for are taken in the interval $\left[sup\left(b_{inf,i}, 0\right), inf\left(b_{sup,i}, \sqrt{q} - 1\right)\right]$ instead of the interval $[b_{inf,i}, b_{sup,i}]$ computed in (13).

In Fig.8, we plot the complexity of this modified SB-Stack decoder for a $4 \times 4$ MIMO system using a $16 - QAM$ constellation. As we can see, this latter exhibits a complexity reduction of at least 30% compared to the sphere decoder while maintaining the ML performances.

### C. Sub-optimal SB-Stack decoder

The SB-Stack algorithm proposed above searches for the shortest path $(x_n, x_{n-1}, \ldots, x_1)$ that minimizes the metric $f(x^{(n)})$. The solution to this problem is then the ML one. However, by introducing a parameter in the cost function as defined in [3], we can rewrite (10) as

$$f(x^{(k)}) = \sum_{i=n-k+1}^{n} f_i(x_i) - b \cdot k \tag{17}$$

where $b \in \mathbb{R}^+$ is called the bias. Each path found in the tree is then weighted with a negative value $-b \cdot k$. Under this constraint, the algorithm advantages the paths of larger lengths since the smallest metrics correspond to the deepest paths in the tree. Hence, this parametrized version allows the SB-Stack decoder to visit less nodes. Consequently the complexity is reduced and the decoder converges more rapidly, however the solution is not guaranteed to be ML but depends on the value of $b$.

In Fig.9.a, we plot the performances defined as the symbol error rate as a function of the SNR, for a $4 \times 4$ MIMO system for different values of the bias. In Fig.9.b, we plot the correspondent complexities.

We show that the performance decreases as $b$ increases, and for a high value of $b$, it approaches the ZF-DFE. In fact, since the decoding rule is no longer the minimization of the euclidean distance as in (10), the performances obtained are not optimal. Though, at small values of $b$, the second term $-b \cdot k$ is negligible in equation (17), the cost function is approximately equal to (10) and then near-ML performances are achieved. The complexity however decreases continuously with $b$.

This parametrized version of the SB-Stack decoder is therefore very interesting since it allows to obtain many ranges of performances with reduced complexities by only changing the value of the bias.

## Part III: Soft decoding using the stack algorithm

In part II, lattice decoding for MIMO schemes is considered and a less complex stack decoder is proposed. This one is presented in the case of hard outputs. However, when a channel coding is integrated to the transmission scheme, the soft decoding becomes necessary. In this part, soft output detection is then investigated and we show that the proposed SB-Stack decoder can be extended to support soft-outputs.

### 1. Overview of Soft decoding

In the literature, soft-output MIMO decoding was studied. Some solutions to this issue have been proposed in [21],[22] and the so called 'list' or 'candidate list' was introduced. The most known soft-output lattice decoder for MIMO systems is the List Sphere Decoder (LSD).

We are interested here in the proposed stack decoder with spherical bounds, which offers an interesting structure and a great advantage to provide a selected list.

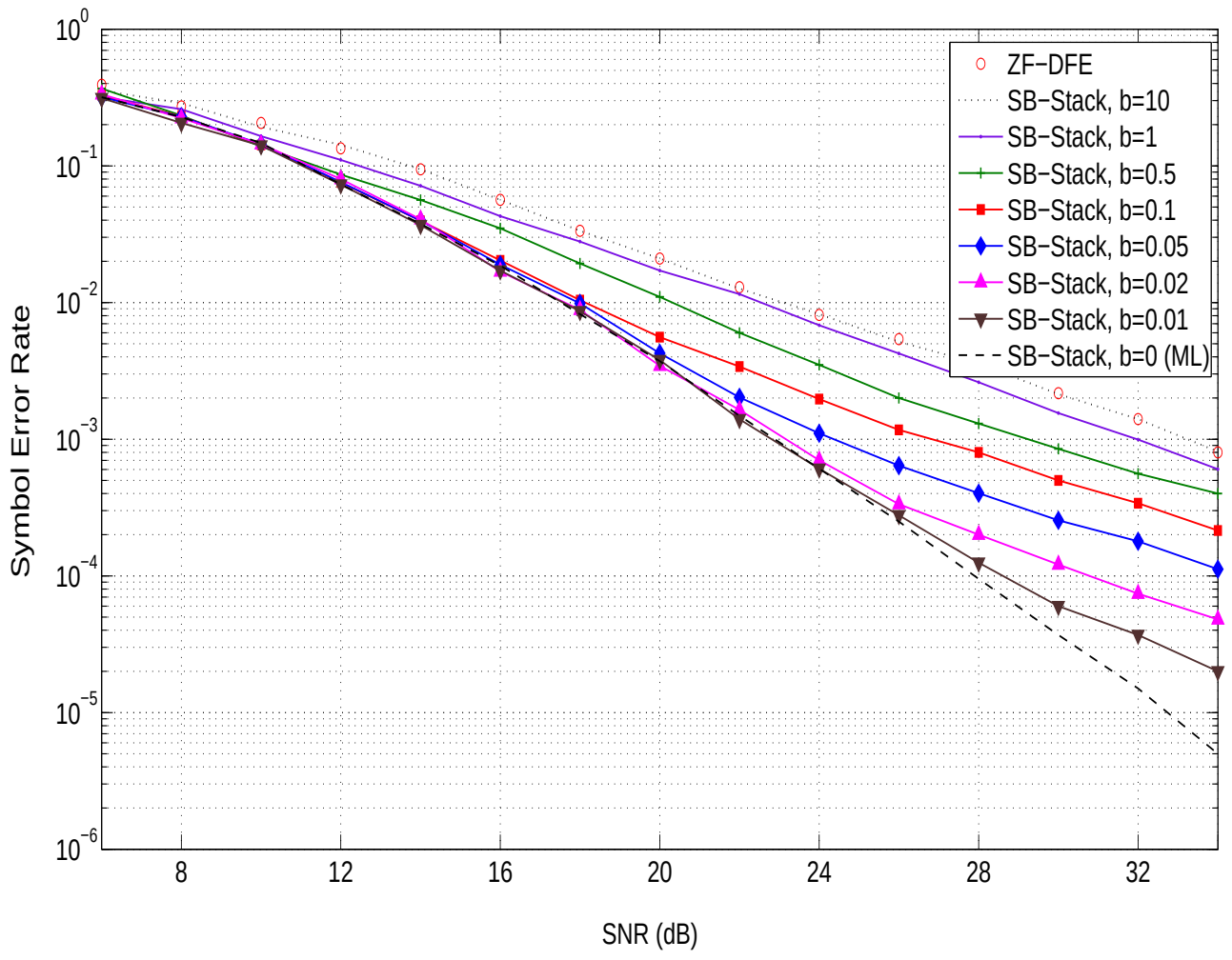

(a)

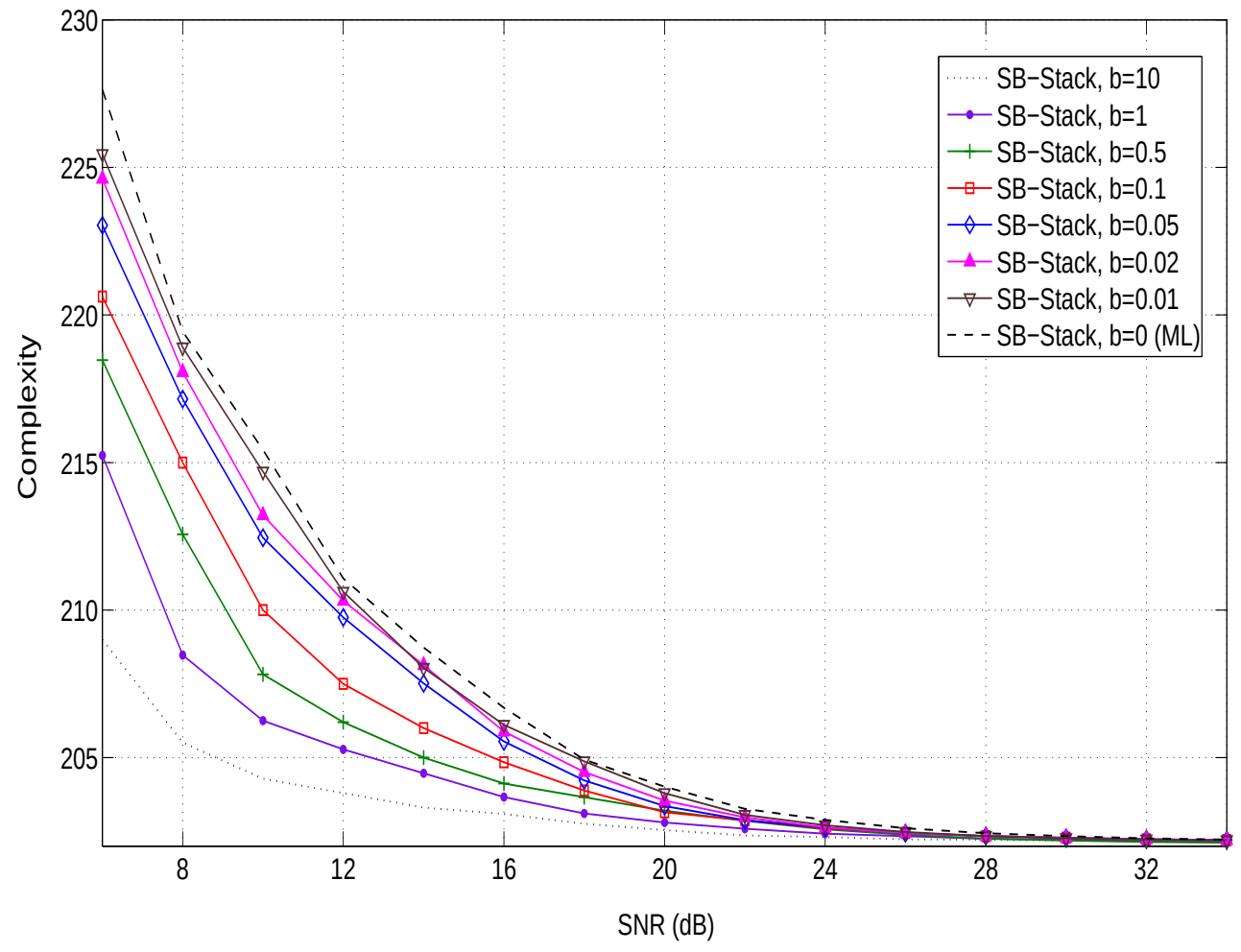

(b)

Fig. 9. Performance and complexity of the SB-Stack decoder for a $2 \times 2$ MIMO system with SM using a $16-QAM$ constellation, for different values of the bias

Soft decoding can be realized using a posteriori probability techniques. A posteriori probability (APP) techniques are a judicious choice for high performance receivers with reasonable complexity. Maximizing the APP for a given bit minimizes the probability of making an error on that bit. The APP is usually expressed as a log-likelihood ratio (LLR) value. A decision is made from a LLR value by using its sign to tell whether the bit is one or zero. The magnitude of the LLR value indicates the reliability of the decision. LLR values near zero correspond to unreliable bits. In the following, the logical zero for a bit is represented by amplitude level $b_k = -1$ and logical one by $b_k = +1$. The modulator maps each layer of the bits into data symbols through the mapping

$$f : \{-1, +1\}^{1\times B} \rightarrow \mathfrak{C}$$

where $\mathfrak{C}$ denotes the data symbol constellation and $B = \log_2 |\mathfrak{C}|$ is the number of bits represented by each data symbol. Let's $K$ denotes the number of symbols belonging to each codeword transmitted in each channel use. The LLR of the $i^{th}$ bit, where $i \in [1, BK]$, is defined as

$$LLR(b_i) = \log \frac{\Pr\left(b_i = +1/\boldsymbol{y}, \boldsymbol{H}\right)}{\Pr\left(b_i = -1/\boldsymbol{y}, \boldsymbol{H}\right)} \tag{18}$$

One can assume equal probability for each data bits (an interleaver at the encoder can be used to scramble the bits). Using Bayes theorem, the bit metric can be written as

$$LLR(b_i) = \log \frac{\sum_{\mathbf{b}\in\mathcal{D}_{i,+1}} \Pr\left(\boldsymbol{y}/\boldsymbol{b}, \boldsymbol{H}\right)}{\sum_{\mathbf{b}\in\mathcal{D}_{i,-1}} \Pr\left(\boldsymbol{y}/\boldsymbol{b}, \boldsymbol{H}\right)}. \tag{19}$$

where $\mathcal{D}_{i,+1}$ and $\mathcal{D}_{i,-1}$ are the set of $2^{BK-1}$ bit vectors $\boldsymbol{b}$ with $b_i$ being $+1$ and $-1$. If we assume an additive zero mean white circularly symmetric complex Gaussian noise, the equation (19) can be written as

$$LLR(b_i) = \log \frac{\sum_{\boldsymbol{b}\in\mathcal{D}_{i,+1}} e^{-\frac{1}{\sigma^2}\|\boldsymbol{y}-\boldsymbol{H}\cdot\boldsymbol{x}(\boldsymbol{b})\|^2}}{\sum_{\boldsymbol{b}\in\mathcal{D}_{i,-1}} e^{-\frac{1}{\sigma^2}\|\boldsymbol{y}-\boldsymbol{H}\cdot\boldsymbol{x}(\boldsymbol{b})\|^2}}. \tag{20}$$

In order to reduce the corresponding computational complexity, one can employ the *max-log* approximation [23] to get

$$\begin{aligned} LLR(b_i) &\approx \max_{\boldsymbol{b}\in\mathcal{D}_{i,+1}} \left\{-\frac{1}{\sigma^2}\left\|\boldsymbol{y} - \boldsymbol{H}\cdot\boldsymbol{x}\left(\boldsymbol{b}\right)\right\|^2\right\} - \max_{\boldsymbol{b}\in\mathcal{D}_{i,-1}} \left\{-\frac{1}{\sigma^2}\left\|\boldsymbol{y} - \boldsymbol{H}\cdot\boldsymbol{x}\left(\boldsymbol{b}\right)\right\|^2\right\} \\ &= \frac{1}{\sigma^2}\left[\min_{\boldsymbol{b}\in\mathcal{D}_{i,-1}} \left\|\boldsymbol{y} - \boldsymbol{H}\cdot\boldsymbol{x}\left(\boldsymbol{b}\right)\right\|^2 - \min_{\boldsymbol{b}\in\mathcal{D}_{i,+1}} \left\|\boldsymbol{y} - \boldsymbol{H}\cdot\boldsymbol{x}\left(\boldsymbol{b}\right)\right\|^2\right] \end{aligned} \tag{21}$$

Soft-output detection on MIMO channels can be achieved via an exhaustive list as in [24] or a limited size list of spherical shape as in [21] and [22].

The APP detector based on an exhaustive has a relatively large complexity exponential in the number of transmit antennas and the number of bits per modulated symbol. In other hand, a non-exhaustive list APP detector is sub-optimal but has a low complexity which is proportional to the list size. Several list decoders were already proposed. We recall in the following the most known ones and we propose a new soft-decoder based on the SB-Stack.

### *A. List Sphere Decoder (LSD)*

An exhaustive search needs to examine all the constellation points. the sphere decoder avoids an exhaustive search by examining only the points that lie inside a sphere with a given radius $C$. The performance of the algorithm is closely tied to the choice of the initial radius $C$. If $C$ is chosen too small, the algorithm could fail to find any point inside the sphere, requiring that $C$ be increased. However, the larger $C$ is chosen, the larger the search will spend time. In [22], a simple modification to the sphere decoder was introduced. The sphere decoder generates a list $\mathcal{L}$ of $N_p$ points. These points make $||\mathbf{y}-\mathbf{Hx}||^2$ smallest among all the points inside the sphere. The list, by definition, must include the ML point. To create $\mathcal{L}$, the sphere decoder needs to be modified in two ways: when a candidate is found inside the sphere, the radius $C$ should not be reduced. In addition, the candidate is added to the list if one of the following conditions is satisfied: either the list is not full or at least one candidate in the list has a higher cost than the new candidate. In this last case, the new candidate replaces the one having the large euclidean distance to the received point. Thus, the constructed list contains the ML point and $N_p - 1$ neighbors for which the square error is smallest. The soft information about any given bit $b_k$ is essentially contained in $\mathcal{L}$: if there are more entries in $\mathcal{L}$ with $b_k = 1$ than those with $b_k = -1$, then it can be concluded that the likely value for $b_k$ is $+1$, whereas if there are fewer entries in $\mathcal{L}$ with $b_k = 1$, then the likely value is $-1$. A larger radius $C$ generally allows for a larger $N_p$ which makes the list more reliable.

There is also a trade-off between the accuracy and the decoding delay of the LSD. Finding $N_p$ points is generally slower than just finding one point, because the search radius always stays fixed and does not decrease with each found point. One problem of this algorithm is the variable size of the list. In[22], a radius, function of the desired number of points, is proposed. The decoding error can be written as

$$\|\boldsymbol{y} - \boldsymbol{H} \cdot \boldsymbol{x}\|^2 = \|\boldsymbol{w}\|^2 \ \ \sigma^2 \chi^2_{2N_p}, \tag{22}$$

where $\chi^2_{2N_p}$ is a chi-square random variable with $2N_p$ degrees of freedom. The expected value of this

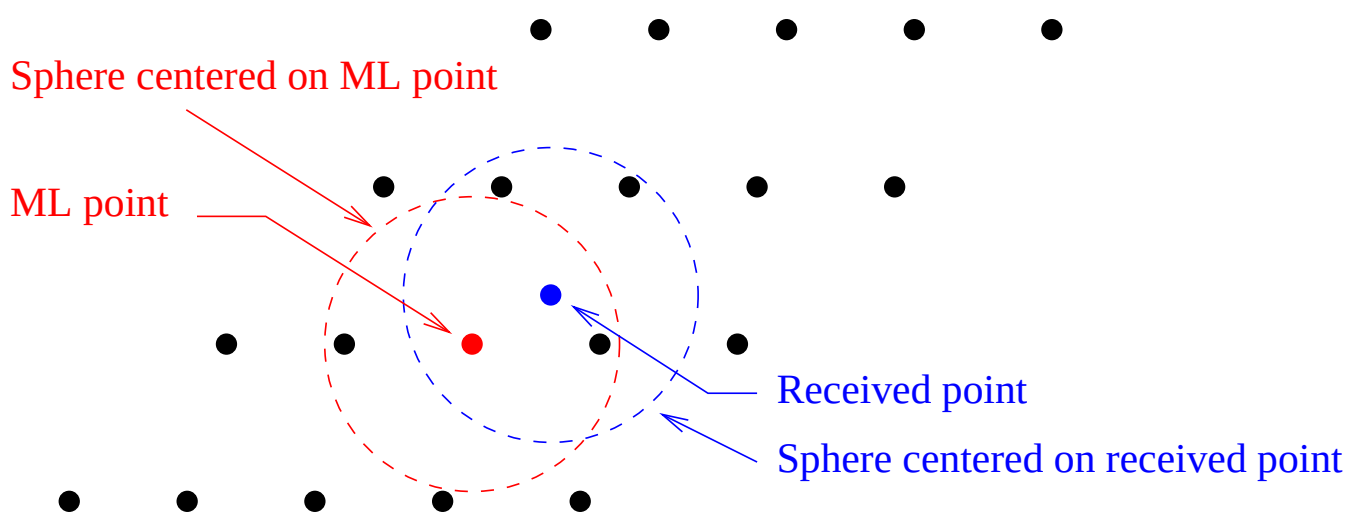

Fig. 10. Sphere centered on the ML point and the sphere centered on the received point

random variable is $\sigma^2 E[\chi^2_{2N_p}] = 2\sigma^2 N_p$. One possible choice of the radius is

$$C^2 = 2\sigma^2 \zeta N_p - y^\dagger \left( I - H \left( H^\dagger H \right)^{-1} H^\dagger \right) y \tag{23}$$

where $\zeta > 1$ is chosen so that one can be reasonably sure, as measured by a confidence interval for the $\chi^2_{2N_p}$ random variable, that the true transmitted $\boldsymbol{x}$ will be found.

The important weak point in the LSD is the instability of the list size. The number of visited points before reaching the ML point can not be fixed exactly, only an approximate number can be provided. The sphere radius is selected to give nearly the needed number. Moreover, the constructed list is not centered at the ML point. A Shifted Spherical List Decoder was proposed in [21] to resolve this problem.

### *B. Shifted Spherical Decoder (SSD)*

The APP detector starts by applying a sphere decoder to find the ML point, then a spherical list centered around the ML point is built. This list depends on the ML point position and the channel state. The trick behind this idea is to center the spherical list $\mathcal{L}$ on the ML point instead of the ZF point. Fig.10 shows in two dimensional lattice the sphere centered on the ML point compared to the one centered on the ZF point.

Usually the received point $\boldsymbol{y}$ is outside the constellation, specially when considering large dimensions $n$. The sphere decoder centered on the received point visits a lot of lattice points to find a small number of constellation points. However, when the sphere is centered on the ML point, the number of enumerated points is reduced and higher likelihood constellation points are considered. But to guarantee a high stability for the number of points required in the list, one should be careful for the choice of the shifted list radius. This radius should take into account the number of points to create the list. In [22], an approximation is made: the volume of the sphere containing $N_p$ points is equal to the volume of $N_p$ fundamental parallelotopes. As a result, the radius $C$ can be computed as

$$C \approx \left( \frac{N \times vol(\Lambda)}{V} \right)^{\frac{1}{n}}, \tag{24}$$

where $vol(\Lambda) = |det(\boldsymbol{H})|$, $\boldsymbol{H}$ is the lattice generator matrix, $n$ is the dimension of $\boldsymbol{H}$ and $V$ is the unit radius sphere volume in the real space $R^n$, $V = \frac{\pi^M}{M}$. This method has the disadvantage of being stable only for high values of $N_p$. If we assume $N_0$ the effective number of points found inside the list $\mathcal{L}$, one can check that

$$\lim_{N_p \to \infty} \frac{N_0}{N_p} = 1 \tag{25}$$

But when considering a finite constellation, $N_0$ will decrease because of the limited shape of the intersection between the sphere and the constellation. This depends on the ML point position inside the constellation and the shape of this constellation. As a result, the radius $C$ of the shifted spherical list for the constellation can be given by

$$C = \left( \frac{\alpha[n_{hyp}] \times \mu_\gamma \times N_p \times vol(\Lambda)}{V} \right)^{\frac{1}{n}}, \tag{26}$$

$\alpha$ is an expansion factor of the list size which depends on the number of hyperplanes $n_{hyp}$ at the constellation boundaries passing through the ML point. $\mu_\gamma$ is an additional expansion factor depending on the shape of the constellation [21].

## 2. Soft Decoding using the Stack Decoder With Spherical Bound (Soft SB-Stack)

### *A. Principle*

We propose here an extension of the classical stack decoder and the new proposed SB-Stack decoder to get soft information outputs. We have modified this algorithm to generate soft-output information in the form of LLR. Stack decoders have the capability of generating a candidate list in their original algorithm. In each iteration, children nodes are generated and stored in the stack ordered in function of their costs. At the end of the algorithm, the first leaf node reaching the top of the stack is the ML point. In this work, we improve the SB-Stack algorithm to make it suitable for a soft output by constructing a list instead of only the ML point. In fact, after the end of the process, one can remark that stack is still full of nodes with different sizes (with different levels in the tree) and no one among them is reaching the top of the stack. The most straightforward idea is to extract the ML point from the original stack, put it in another stack and continue the searching phase. The next node reaching the top of the stack is also removed and putted in the second stack with its corresponding cost. There is two possibilities to stop the algorithm:

- either we fix the number of points in the list (the size of the second stack). In this case the algorithm continues in this manner until the second stack will be full.
- another possible criterion is to fix a lower bound on the node costs (worst cost to be admissible), and when the cost falls below this limit value, the algorithm gives up.

Thus, only the nodes stored in the second stack will contribute to the soft decision. This new Soft SB-Stack decoder is an extension of the first one and aims at generating more leaf nodes. The second stack is used later to generate the LLR. The main advantages of this algorithm are

- the stability: the algorithm will stop as soon as the number of candidates is reached. The issue regarding the computation or estimation of the ideal radius value is removed.
- the list is centered at the ML point. In other words, the list is filled up with the closest points only in an ascending cost order, leading to an optimal LLR computation for a given list size.
- a low complexity since we only pursue the stack algorithm with no additional search method and exploit the nodes being already computed and still in the stack.

The disadvantage of all the previous soft decoders is their inability to provide soft outputs with low complexity, and the worst case corresponds to the exhaustive search.

The Soft SB-Stack decoder provides less complexity than these latters. Moreover, we can apply the bias parameter as in the equation (17). This leads to a trade-off of performance-complexity with soft outputs, which was never done before. In the case of a non-null bias, the ML point is reached later and belongs to the list. One can even impose an aggregate run time constraint.

A straightforward conclusion is the flexibility of the new Soft SB-Stack decoder with practical constraints that system engineers can be faced with the design of the receiver.

### *B. Simulations and results*

In this part, we illustrate the application of the new Soft SB-Stack decoder for MIMO space time transmission. The binary information is encoded using a rate R-convolutional code. The coded bits are fed to a q-QAM mapper (Gray mapping) that generates symbols. The spectral efficiency in the V-BLAST case is $R \times B \times M$ bits per channel use, $B = \log_2(q)$.

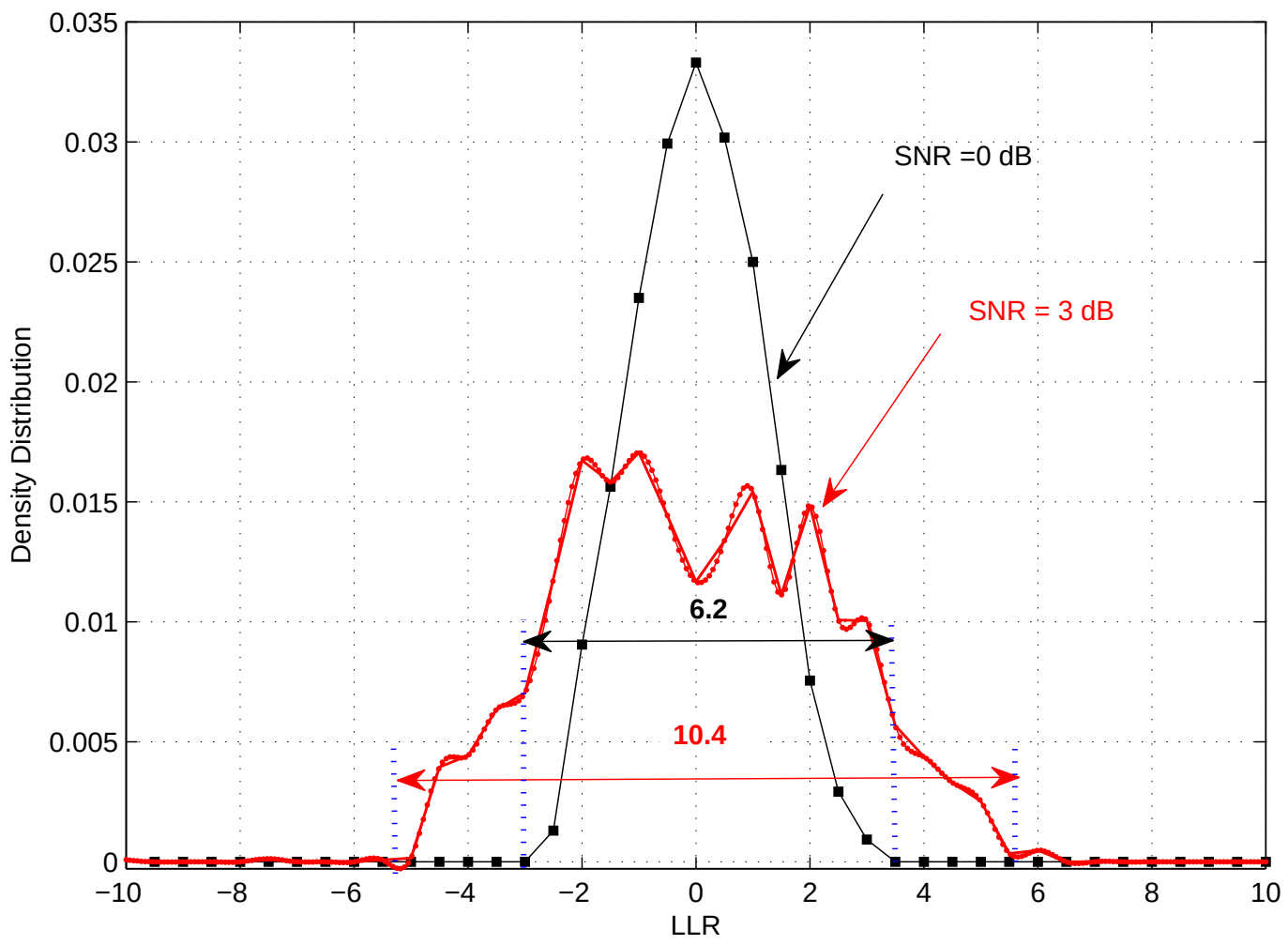

Fig. 11. LLR Density Distribution for SNR=0 dB and SNR=3 dB

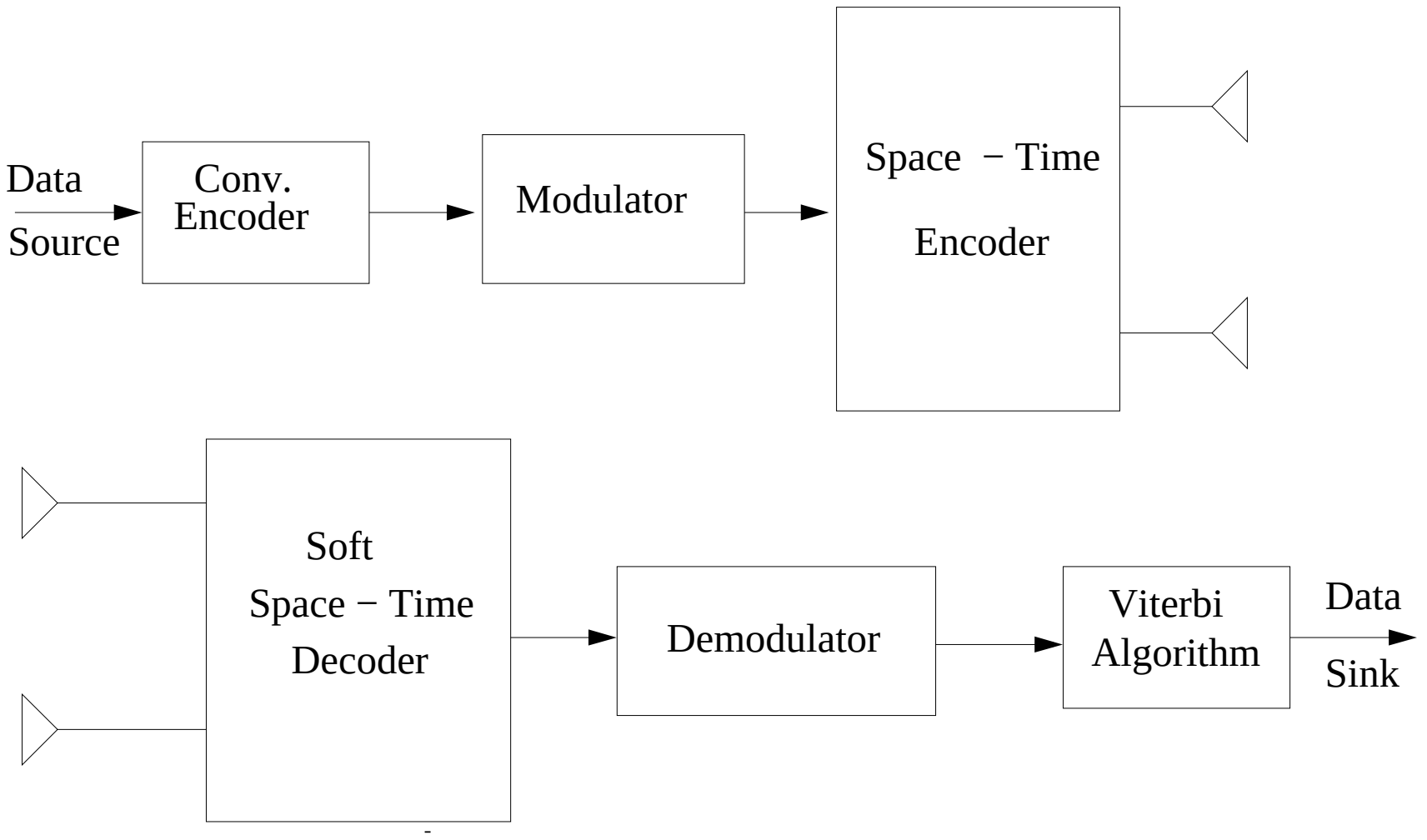

Fig. 12. Diagram of transmitter and receiver with soft space time decoding

Fig.11 shows the LLR distribution of the candidates found inside the stack for $SNR = 0dB$ and $SNR = 3dB$. It can be observed that when SNR increases, the LLR distribution curve is going to get a concave shape with a cavity around zero. This can be expected since $zero - LLR$ means ambiguity in the decision which is diminished when SNR increases.

For high SNR, the LLR values stretch to infinity and in practice they are saturated to a high chosen value. The LLR distribution curves provide us with information about the intervals to which LLR belong.

LLR will be sampled into $2^m - 1$ levels of their interval distribution and then quantized to $m$ bits to serve as input for the soft Viterbi decoder.

Fig.12 shows the diagram of the simulated transmitter and receiver. We consider the decoding of 200 information bits per frame, a rate $\frac{1}{2}$-convolutional code modulated by a $4-QAM$ constellation. The generator $G = (7, 5)$ (in octal notation) has the memory of $T = 3$. Two transmitting and two receiving antennas are used and symbols are spatially multiplexed. In the reception side, for soft input Viterbi decoder, LLR are provided by the space time decoder using a list of candidates.

In Fig.13.a, we plot the BER (Bit Error Rate) as a function of $E_b/N_0$ under a Rayleigh flat fading channel. Monte Carlo simulations are used with $10^6$ distinct channel realizations, where the channel realization remains constant over one signal block and changes randomly from block to block. We assume that the channel state information is available at the receiver.

This figure shows a comparison between the different soft decoders cited above. In this simulation, the size of the candidate list of the unconstrained LSD (ULSD) is not predetermined as in the original LSD. One can fix a sphere radius which determines the average size of the candidate list which is not constrained, therefore it is referred to this as Unconstrained List Sphere Decoder (ULSD) [25].

For the SSD [21] and the Soft SB-Stack decoders, we take a list of 6 candidates. We can see that he Soft SB-Stack decoder outperforms the other decoders in term of performance and exhibits a gain over 1dB compared to the SSD. The achieved improvement is up to 2 dB compared to the LSD. This is due to the fact that the SB-Stack decoder is more flexible for increasing the list size and the algorithm can continue running to get more candidates which is not the case of the ULSD and the SSD which are constrained by the chosen radius.

Next, we compare the complexity of the different soft decoders. In Fig.13.b, we plot the number of multiplications needed to decode one transmitted codeword. As might be expected, the Soft SB-Stack decoder enjoys an advantageous average complexity compared to the SSD. However, it is outperformed by the LSD in term of complexity but the performance of this later is worse. The improved performance of the SB-Stack decoder compared to the LSD comes at a cost of additional complexity requirements.

## CONCLUSION

The main purpose of this work was to apply sequential decoders, especially the stack one, for multi-antenna systems over linear Gaussian channels. We saw that the MIMO decoding can be tasked on a CLPS problem. Toward this end, we proposed approaches to apply the stack decoder to decode lattice. Our main contribution was to introduce a novel version of the stack decoder for MIMO systems with a reduced complexity. The proposed decoder, that we called the Spherical Bound Stack decoder, consists on a modified stack decoder combining both the stack algorithm search strategy and the sphere decoder properties. In a first time, this modified decoder was introduced to decode lattice. In a second time, we brought the necessary modifications to apply it in the case of finite constellations and we showed by simulations that the SB-Stack decoder outperforms the classical lattice and MIMO ones in term of complexity.

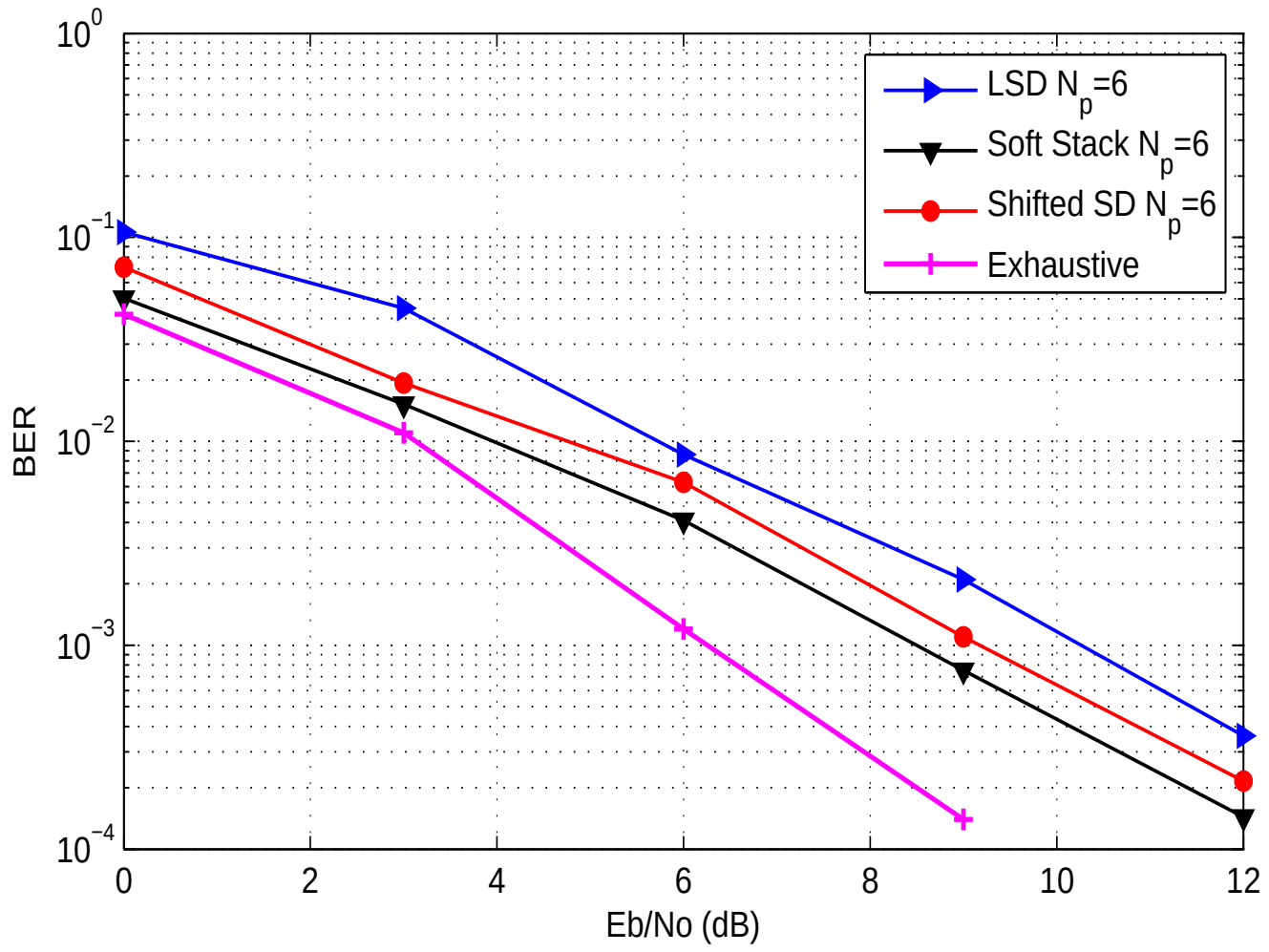

(a)

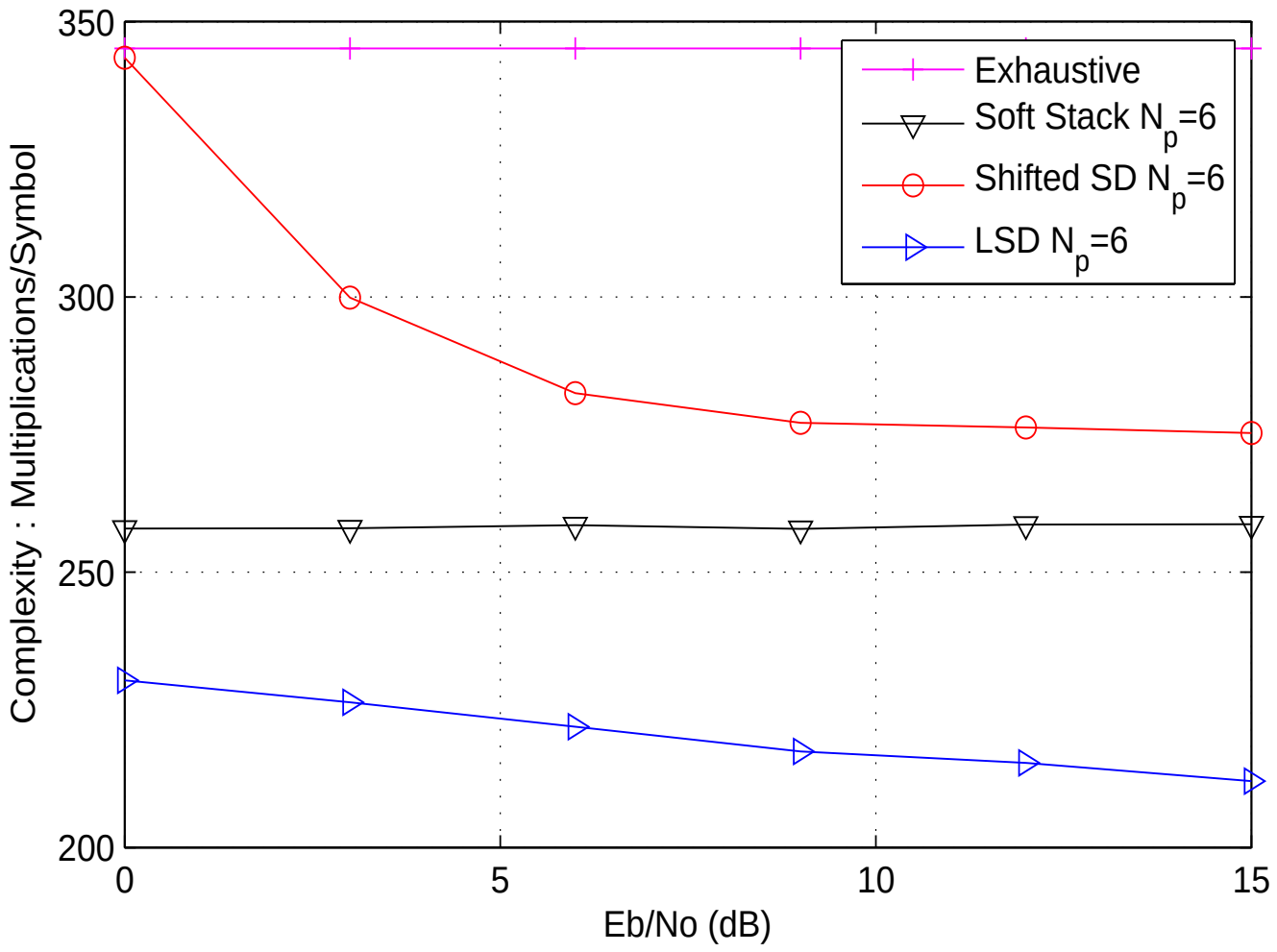

(b)

Fig. 13. Performance on $2 \times 2$ ergodic MIMO with $4-QAM$, rate $\frac{1}{2}$ convolutional encoder

As a second part, we extended the SB-Stack decoder to support soft output MIMO detection. By exploiting the advantage of the memory use to deliver a soft output, a good improvement in the performances is distinguished. The simulation results show that the Soft SB-Stack decoder outperforms the known List Sphere Decoder and the Shifted Sphere Decoder and is at 2dB from the ML performances. Moreover, due to the stack properties, the Soft SB-Stack decoder is also easily implemented and is more flexible for an increase of candidates list size.